\documentstyle[prd,epsfig,aps,floats]{revtex}

\newcommand{\Sslash}[1]{ \parbox[b]{0.6em}{$#1$} \hspace{-0.55em}
                         \parbox[b]{0.55em}{ \raisebox{-0.2ex}{$/$}}}
\newcommand{\beq}{\begin{equation}}
\newcommand{\eeq}{\end{equation}}
\newcommand{\beqa}{\begin{eqnarray}}
\newcommand{\eeqa}{\end{eqnarray}}

\newcommand{\lsim}{\lesssim}
\newcommand{\gsim}{\gtrsim}
\newcommand{\half}{\frac{1}{2}}
\newcommand{\shalf}{{\textstyle \frac{1}{2}}}

\newcommand{\ie}{{\it ie.}}
\newcommand{\eg}{{\it eg.}}
\newcommand{\cf}{{\it cf.}}
\newcommand{\etal}{{\it et al.}}

\newcommand{\M}{{\cal M}}
\newcommand{\Ps}{{\cal P}}
\newcommand{\R}{{\cal R}}

\newcommand{\jpsi}{J/\psi}
\newcommand{\order}[1]{${\cal O}(#1)$}
\newcommand{\morder}[1]{{\cal O}(#1)}
\newcommand{\eq}[1]{Eq.\ (\ref{#1})}

\newcommand{\ptr}{p_\perp}

\newcommand{\llb}{{\lambda\bar\lambda}}
\newcommand{\ssb}{{\sigma\bar\sigma}}

\newcommand{\pvec}{\bbox{p}}
\newcommand{\Pvec}{\bbox{P}}
\newcommand{\kvec}{\bbox{k}}
\newcommand{\qvec}{\bbox{q}}
\newcommand{\epsvec}{\bbox{\varepsilon}}
\newcommand{\evec}{\bbox{e}}
\newcommand{\lvec}{\bbox{\ell}}
\newcommand{\xvec}{\bbox{x}}

\newcommand{\as}{\alpha_s}

\newcommand{\qpair}{Q\bar Q}
\newcommand{\cpair}{c\bar c}

\newcommand{\PL}[3]{Phys.\ Lett.\ {{\bf#1}}, {#2} ({#3})}
\newcommand{\NP}[3]{Nucl.\ Phys.\ {{\bf#1}}, {#2} ({#3})}
\newcommand{\PR}[3]{Phys.\ Rev.\  {{\bf#1}}, {#2} ({#3})}
\newcommand{\PRL}[3]{Phys.\ Rev.\ Lett.\ {{\bf#1}}, {#2} ({#3})}
\newcommand{\ZP}[3]{Z. Phys.\ {{\bf#1}}, {#2} ({#3})}

\begin{document}

\twocolumn[\hsize\textwidth\columnwidth\hsize\csname @twocolumnfalse\endcsname
\title{%
\hbox to\hsize{\normalsize\hfil\rm NORDITA-98/43 HE}
\hbox to\hsize{\normalsize\hfil hep-ph/9806424}
\hbox to\hsize{\normalsize\hfil \protect\today}
\vskip 40pt
Quarkonium Production through Hard Comover Scattering\cite{byline1}}
\author{Paul Hoyer and St\'ephane Peign\'e\cite{byline2}}
\address{Nordita\\
Blegdamsvej 17, DK--2100 Copenhagen, Denmark\\
www.nordita.dk}

\maketitle

\begin{abstract}
We propose a qualitatively new mechanism for quarkonium production,
motivated by the global features of the experimental data and by the
successes/failures of existing models. In QCD, heavy quarks are created
in conjunction with a bremsstrahlung color field emitted by the colliding
partons. We study the effects of perturbative gluon exchange between the
quark pair and a comoving color field. Such scattering can flip the spin
and color of the quarks to create a non-vanishing overlap with the wave
function of physical quarkonium. Several observed features that are
difficult to understand in current models find simple explanations.
Transverse gluon exchange produces unpolarized $\jpsi$'s, the $\chi_{c1}$
and $\chi_{c2}$ states are produced at similar rates, and the anomalous
dependence of the $\jpsi$ cross section on the nuclear target size can be
qualitatively understood.
\end{abstract}
\vskip2.0pc]


\section{Introduction and Summary} \label{sec1}

\subsection{Introduction} \label{sec11}

Quarkonium production has turned out to be a challenge as well as an
inspiration for our understanding of hard QCD processes
\cite{rev1,rev2,rev3,rev4,rev5,rev6,rev7}. In the case of standard
inclusive processes, the theoretical framework is uniquely defined by
the QCD factorization theorem \cite{fact}. This theorem allows a
physical cross section $\sigma$ to be expressed as a product of universal
parton distribution and fragmentation functions multiplied by a
subprocess cross section $\hat\sigma$, which is calculable in PQCD. The
factorization theorem relies on a completeness sum over the final state
and does not apply to the quarkonium cross section, which constitutes
only a small fraction of the total heavy quark production cross section.

While a theoretical description of quarkonium production is thus more
model dependent, it can potentially reveal more about the dynamics of
hard processes than can be learned from, \eg, the total heavy quark cross
section. In particular, it is intuitively plausible that the quarkonium
cross section is sensitive to reinteractions with partons
created along with the heavy quark pair. Thus quarkonium production can
serve as a `thermometer' of the environment, as has been recognized in
the search for a quark-gluon plasma in heavy ion collisions \cite{masa}.
In this paper we wish to explore the possibility that rescattering of the
heavy quarks causes the puzzling anomalies seen in quarkonium
hadroproduction.

There is independent evidence that the environment in charm
hadroproduction is rather `hot'. In $\pi^- N \to D\bar D + X$ the
observed spread in relative azimuthal angle of the $D$-mesons requires an
average intrinsic transverse momentum of the incoming partons $\langle
k_\perp^2 \rangle \simeq 1$ GeV$^2$ \cite{frixione}. The `leading
particle' asymmetry between $D^-$ and $D^+$ is larger than expected
from PQCD, and persists for $D$-mesons produced with $k_\perp^2 \lsim 10$
GeV$^2$ \cite{asymhad}. Both effects are weaker for photoproduced charm
\cite{frixione,asymphoto}.

We shall study the effects of perturbative gluon exchange between the
heavy quark pair and a comoving \cite{bmvg} color field. The interaction is
assumed to occur at an early stage, before the pair has expanded to the size of
physical quarkonium. Hence only comovers which are created (via bremsstrahlung)
in the hard process itself are relevant, whereas interactions with beam
fragments at typical hadronic distances $\sim 1$ fm are ignored. In this sense
our approach differs from that of the `Color Evaporation Model' (CEM)
\cite{cem1,cem2}, which only considers late, non-perturbative interactions of
the heavy quarks. On the other hand, similarly to the CEM our quark
pairs are produced near threshold. Hence many of the phenomenological
successes of the CEM concerning the dependence of quarkonium cross sections on
various kinematical variables are incorporated in our model. The `Color Octet
Model' (COM) \cite{com}, also considers late interactions, through an
expansion in powers of the relative velocity $v$ of the bound quarks as
specified by NRQCD \cite{nrqcd}. This expansion is general and should hold for
any description of quarkonium production, including ours. The higher order
$(v/c)^n$ terms need not, however, give a dominant contribution to the cross
section. To our knowledge, the COM assumption that the heavy quark pair is
unaffected by earlier reinteractions with its environment has not been proven.

Data on charmonium and bottomonium production is available for a
wide variety of beams, targets and kinematical conditions. Comparisons
with the COM and CEM approaches have met with some successes, but also
with difficulties \cite{rev1,rev2,rev3,rev4,rev5,rev6,rev7}. The data
suggests a production dynamics which in some respects differs from the
late and soft reinteraction scheme of the CEM and COM. In particular,

\begin{itemize}
\item[(A)] The heavy quark pair turns color singlet at an early stage,
while the pair is still compact (\ie, small compared to the size of the
quarkonium wave function).
\item[(B)] In hadroproduction there is at least one secondary gluon exchange
after the primary, heavy quark production vertex.
\item[(C)] The `anomalies' of quarkonium production depend only
weakly on the quark mass $m$, on the CM energy and on the transverse
momentum $p_{\perp}$.
\end{itemize}

We discuss the experimental basis for these features in Section~\ref{sec2}
and then develop our QCD scenario in Section~\ref{sec3}.
This scenario applies to quarkonium production at both moderate and large
transverse momentum. However, we shall limit our discussion and the
calculations presented in Section~\ref{sec4} to the total quarkonium
cross section, \ie, moderate $x_F$ and $p_{\perp} \sim m$. In the rest of
the present Section we summarize our results. Conclusions and an outlook
are presented in Section~\ref{sec5}.

\subsection{Summary} \label{sec12}

The basic Color Singlet Mechanism (CSM) \cite{csm}, which is known to grossly
underestimate the $\jpsi$ hadroproduction cross section, is shown
in Fig.~1a. The gluon emission takes place in a (proper) time $\tau\sim 1/m$,
simultaneously with the heavy quark production process. This
is compatible with feature (A), but since there are no relevant later
interactions the CSM does not agree with (B). The situation is
qualitatively the same for loop corrections to the CSM (Fig.~1b), since
the space-time scale of the loop is $1/m$.

Prior to the heavy quark production vertex the colliding partons radiate
gluons as part of the normal QCD structure function evolution. The
space-time scale of this process is determined by the virtuality $k^2$ of
the partons which couple to the heavy quark line. As is characteristic of
evolution processes, the $k^2$ distribution $\as(k^2)dk^2/k^2$ is
logarithmic between a lower cutoff determined by the (perturbative) 
factorization scale
and an upper limit given by the heavy quark mass $m$. 
Thus the effective value of $|k^2|$ is given by a {\it perturbative}
scale which we denote by $\mu^2$. This scale grows with $m^2$ but satisfies
$\mu^2 \ll m^2$. We will investigate the effects of rescattering at this
hardness scale $\mu$.
\begin{figure}[h]
\center\leavevmode
\epsfxsize=6.2cm
\epsfbox{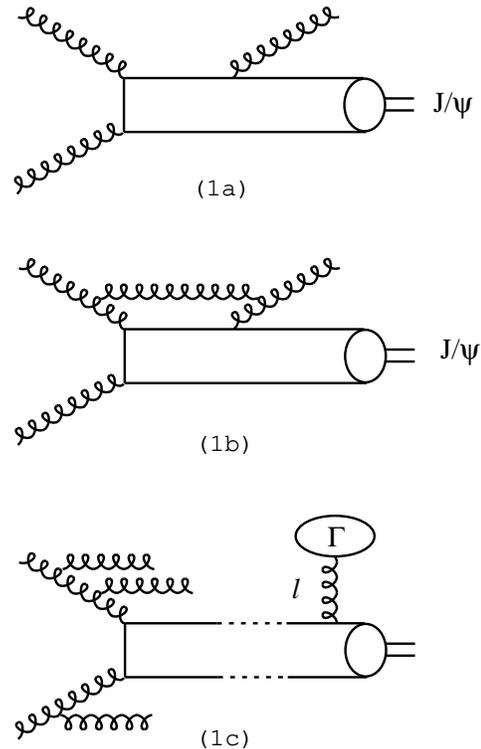}
\medskip
\caption{Basic processes for $\jpsi$ hadroproduction in the CSM (figures
(1a) and (1b)) and in our model (figure (1c)).}
\end{figure}

The approach presented in Section~\ref{sec3} is based on a perturbative
reinteraction of momentum transfer $\ell \sim \mu$ between the heavy
quarks and a classical color field $\Gamma$ (Fig.~1c). This field is
assumed to originate from gluon bremsstrahlung in the $gg\to \qpair$
subprocess. The reinteraction can occur long after the heavy quarks
are created provided the field $\Gamma$ is comoving with the quark
pair. Hence we shall assume $\Gamma$ to be isotropic in the pair rest
frame. The scale $\mu$ of the secondary interaction is smaller than
the quark mass scale $m$ of the CSM but larger than the bound state scale
$\as m$ of the COM\footnote{For the charmonium system, some of these
scales are numerically similar, but should be distinguished for reasons of
principle. The scales do differ for bottomonium.}. The existence of the
comoving color field $\Gamma$ in hadroproduction is our main postulate,
motivated by the data. There should be no corresponding field in the
current fragmentation region of photo- and leptoproduction, since photons
do not radiate gluons (at lowest order).

\begin{figure}[h]
\center\leavevmode
\epsfxsize=8cm
\epsfbox{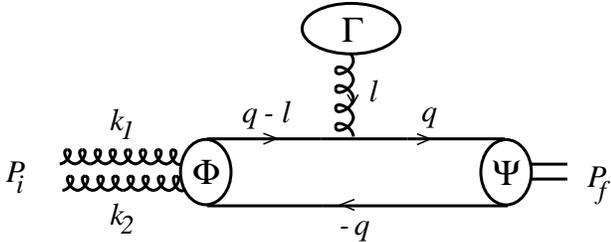}
\medskip
\caption{A perturbative interaction between the quark pair and a gluon
from the color field $\Gamma$ creates an overlap between the $\qpair$
wave function $\Phi$ from the $gg \to \qpair$ process with the physical
quarkonium wave function $\Psi$. There is a second diagram where the gluon
attaches to the antiquark.}
\end{figure}

Our quarkonium hadroproduction amplitude is essentially given by the
perturbative diagram of Fig.~2 (together with the diagram corresponding
to rescattering of the antiquark). Here $\Phi$ is the (color octet) wave
function of the heavy quark pair produced in the fusion process $gg \to
\qpair$ and $\Psi$ is the (color singlet) quarkonium wave function.

As we shall see, there are two production mechanisms for spin triplet
$S$-wave quarkonia such as the $\jpsi$. In the $gg \to \qpair$ subprocess
the quark pair can either be produced in an $S=L=0$ state, followed by a
spin-flip interaction with a transverse gluon from the color field
$\Gamma$, or the pair can be produced with $S=L=1$, followed by a
spin-conserving interaction with a longitudinal gluon. The first
contribution gives unpolarized quarkonia, since the quark pair is
produced with total angular momentum $J=0$ and the color field
$\Gamma$ is isotropic. The second contribution turns out to give
quarkonia with a transverse  polarization. The striking experimental fact
\cite{badier,poln,heinrich} that the $\jpsi$ and $\psi'$ are produced
unpolarized (at moderate $p_{\perp}$) thus implies that the former
mechanism dominates, \ie, that the gluons in $\Gamma$ are transversely
polarized.

The $P$-wave quarkonium states $\chi_J$ are produced from quark pairs with
$S=0$, $L=1$ followed by a spin-flip interaction with a transverse gluon
from $\Gamma$. The calculated cross section ratio $\sigma(\chi_1) /
\sigma(\chi_2) = 3/5$, in agreement with the ratio
measured in $\pi N$ collisions for charmonium, $0.6\pm 0.3$
\cite{e705,e672}. We also find that due to the indirect $\chi_{cJ} \to
\jpsi +\gamma$ contributions the total $\jpsi$ polarization is slightly
longitudinal, $\lambda = -.14$ (\cf\ \eq{lamdef}). Taking
into account also the CSM mechanism, which dominantly produces
$\chi_{c2}$'s with $J_z=\pm 2$ that decay into transversely polarized
$\jpsi$'s
\cite{vhbt}, the expected ratio $\sigma(\chi_1) /
\sigma(\chi_2) < 3/5$ and $\lambda > -.14$. This is compatible with data
as long as the CSM does not dominate the rescattering mechanism (Fig.~2)
for $\chi_2$ production.

The $\chi_J$ wave function $\Psi$ (Fig. 2) vanishes at the origin,
which suppresses its overlap with small sized $(\sim 1/m)$ heavy quark
pairs. Thus in the CSM the relative production rate of $P$- and
$S$-wave charmonium states is governed by the small ratio
$|R{_\chi}'(0)/2m|^2/|R_\psi(0)|^2 \simeq 0.01$ \cite{eiqu}. In our
approach the initially compact quark pair expands, with velocity
$\sim\mu /m$, before the gluon exchange in Fig. 2 gives the
quark pair the bound state quantum numbers. The wave function
$\Phi$ is thus an expanded version of the quark pair wave
function created in the $gg \to \qpair$ subprocess. We model this by
a scale factor $\rho$, which we fit to the measured
$\sigma(\chi_{cJ}) /\sigma_{dir}(\jpsi)$ cross section ratio. We find
that we need $\rho \simeq 3$, suggesting significant expansion of
the quark pair before reinteraction in $\chi_{cJ}$ production.

Our approach can also be applied to quarkonium production at high
$\ptr \gg m$, where the dominant production mechanism is gluon
fragmentation \cite{bryu}. The fragmenting gluon is initially highly
virtual and radiates gluons with hardness ranging from
the factorization scale up to $\ptr$.
The gluons of relatively small hardness $\sim \mu$
can form a color field comoving with the quark pair. 
We plan to study the detailed predictions of our scenario for high 
$\ptr$ quarkonium production in a future publication.

We shall also not discuss here the special features of quarkonium production
which appear at high $x_F$, and may be related to intrinsic charm \cite{ic} and
scattering from light constituents \cite{bhmt}. Thus our discussion is limited
to the bulk of the charmonium cross section only, which (at fixed target
energies) originates from partons with $\langle x \rangle \sim 0.1$ and
$\langle \ptr \rangle \sim m$.


\section{Qualitative Features of the Data} \label{sec2}

The data on quarkonium production shows many interesting features and
regularities. Several of them are left unexplained (some are even
contradicted) by the dynamics assumed in the Color Octet Model (COM) and
the Color Evaporation Model (CEM). Here we wish to make the
phenomenological case for the three general features (A -- C) of the
production dynamics that we listed in Section~\ref{sec1}.

\subsection{Early Color Neutralization} \label{sec21}

Heavy quarks are produced in $\qpair$ pairs of (transverse) size $\sim
1/m$, where $m$ is the quark mass. The pair is thus initially much smaller
than the Bohr radius of quarkonium bound states, which is of order $1/(\as
m)$. If the quark pair ceases to interact with its surroundings (in
particular, turns color singlet) while it is still in such a compact
configuration then the production rates of all $nS$ states are
proportional to $|R_n(0)|^2$, the square of their wave functions at the
origin. Analogous proportionality holds for the other $^{2S+1}L_J$
quarkonium states.

The above argument requires {\em no} assumption about how the compact
pair is produced. The `$R(0)$ proportionality test' is thus a good
indication of whether the color neutralization occurs early or late, as
measured by the size of the quark pair. This test should moreover be quite
sensitive since the higher radial excitations have a mass near open
flavor threshold. Late scattering of the $\cpair$ system will thus affect
the $\psi'$ (44 MeV below $D\bar D$ threshold) more than the $\psi$ (630
MeV below threshold). This is supported by the observation that
the $\sigma(\psi')/\sigma(\jpsi)$ ratio is significantly {\em reduced} in
central nucleus-nucleus $(SU)$ collisions \cite{lourenco}, as would be
expected due to late interactions with comoving nuclear fragments (or
plasma).

It has been pointed out \cite{knnz} that due to the moderate mass of the
charm quark the wave function of charmonium is probed beyond its origin.
In particular, since the diffractive $\qpair$ photoproduction amplitude is
proportional to the {\em square} of the transverse $\qpair$ separation,
the overlap integral between the quark pair and the bound state wave
functions gets a negative contribution beyond the first node in the radial
wave function of the $\psi'$. The predicted \cite{knnz} ratio
$\sigma(\psi')/\sigma(\jpsi)= 0.17$ is thus smaller than the $|R(0)|^2$
ratio and agrees with a recent HERA measurement $0.150 \pm 0.027 \pm
0.018 \pm 0.011$ \cite{h1ratio}.

The $\sigma(\psi')/\sigma(\jpsi)$ ratio is remarkably universal in
inelastic hadroproduction processes, being nearly independent of the
nature of the beam hadron and the target nucleus, and also of the energy
and of $x_F$ \cite{vhbt,lourenco,gksssv}. This also holds for
$\Upsilon(nS)$ states. The measured ratio for the directly produced
$\jpsi$ cross section (from which decay contributions have been
subtracted) moreover is consistent with \cite{vhbt}
\beq
\frac{\sigma(\psi')}{\sigma_{dir}(\jpsi)} =
\frac{\Gamma(\psi' \to e^+e^-)} {\Gamma(\jpsi \to e^+e^-)}
\frac{M^3_{\jpsi}}{M^3_{\psi'}} \simeq 0.24
\label{psiratio}
\eeq
The hadroproduction ratio (\ref{psiratio}) is somewhat {\em
larger} than the one measured in diffractive photoproduction, indicating
that in hadroproduction the inelastic cross section is more closely
proportional
to
the wave function at the origin\footnote{In the approach
discussed in this paper, the scattering amplitude is proportional only to
the {\em first} power of the $\qpair$ separation. There is also high
$\ptr$ data from the Tevatron \cite{rev3} which indicates that the
$\sigma(\psi')/\sigma(\jpsi)$ ratio is still larger than in
\eq{psiratio}. This may imply an even more pointlike production
dynamics at large $\ptr$.}.

Interesting subtleties aside, the data clearly suggests
the relevance of the perturbative $\qpair$ wave function for quarkonium
photo- and hadroproduction. In the COM and CEM approaches, on the other
hand, the heavy quark pair turns into a color singlet only after it has
expanded to a size comparable to that of the bound state. There is then no
reason to expect the cross section ratio to satisfy \eq{psiratio}
(although this value is also not excluded by those models). We believe
that the agreement of the quarkonium cross section ratios with
expectations based on the wave function of the quarks created in the hard
subprocess is not an accident. This implies that the pair decouples from
its environment while it is still compact (except in the presence of
nuclear comovers \cite{bmvg}).  

\subsection{Reinteraction with a Color Field} \label{sec22}

Quarkonium data provides two indications that a rescattering of the heavy
quark pair with a comoving color field is important in hadroproduction.

\subsubsection{Photoproduction} \label{sec221}

Photons do not radiate gluons\footnote{Except via higher order resolved
processes which are unimportant here.}. At the early stages of
heavy quark creation through the $\gamma g \to \qpair$ subprocess we
should therefore expect {\em no} comoving color field in the photon
fragmentation region. With the rescattering process of Fig.~1c thus
eliminated, the production process should be dominated by the Color
Singlet Mechanism (CSM) of Fig.~1a. It is indeed one of the remarkable
facts of quarkonium production that the CSM works very well for
inelastic $\jpsi$ photoproduction \cite{kzsz,h1photo,zeusphoto}, whereas
the same model underestimates the hadroproduction cross section by an
order of magnitude \cite{rev2,rev3,rev4}. This suggests that
hadroproduction dynamics is coupled to initial parton bremsstrahlung.

The COM parameters which fit $\jpsi$ hadroproduction tend to
overestimate the photoproduction cross section
\cite{rev7,h1photo,zeusphoto}, although it is possible that the
discrepancy could be due to higher order effects \cite{highcom}.

\subsubsection{Nuclear Target Dependence} \label{sec222}

Cross sections of hard incoherent processes on nuclear targets $A$ are
expected to scale like the atomic number of the target, $\sigma(A) \propto
A^{\alpha}$, with $\alpha \simeq 1$. Modifications due to the
$A$-dependence of the quark structure functions are minor in the
presently relevant kinematic range. This is verified by high mass lepton
pair production (the Drell-Yan process), for which $\alpha \simeq 1.00$
is observed \cite{lourenco}.

Charm quark pairs produced in the beam fragmentation region have
large Lorentz factors and expand only after leaving the nucleus. While
compact, the uncertainty in the energy of the $\cpair$ pairs is large and
they couple both to open charm ($D\bar D,\ \bar D\Lambda_c, \ldots$) and
to quarkonium ($\jpsi,\ \psi',\ldots$) channels. In the absence of
effects due to partons comoving with the pair one should therefore expect
the {\em same} $A$-dependence for open and hidden charm, with $\alpha
\simeq 1$.

Data shows that $\alpha = 1.02\pm .03 \pm .02$ for $D/\bar D$ production
at $\langle x_F \rangle = 0.031$ \cite{dprod}, whereas $\alpha = 0.92\pm
0.01$ for $\jpsi$ and $\psi'$ \cite{lourenco}. The deviation of $\alpha$
from unity for charmonium appears to be independent of $E_{CM}$, and
increases with the charmonium momentum fraction $x_F$. This, and the
similar $A$-dependence of $\jpsi$ and $\psi'$, indicates that the nuclear
suppression is not due to an expansion of the $\cpair$ inside the
target. A plausible explanation is that a further interaction, beyond
the nucleus, between the heavy quark pair and comoving gluons is required
for charmonium formation.

Based on the measured $A$-dependence of charmonium production an effective
'absorption' cross section $\sigma_{abs} \simeq 7.3 \pm 0.6$ mb was
obtained \cite{klns} in a Glauber framework. This cross section is too
large for a compact $\cpair$ pair \cite{lourenco,kopeliovich,kharzeev}
and should, in the present framework, be interpreted as the joint cross
section of the $\cpair$ pair and the comoving gluon field. The size of
$\sigma_{abs}$ is then reasonable, since the gluons are at a
relatively large distance $\sim 1/\mu$ from the quark pair. An
analogous interpretation of the nuclear suppression, albeit in a
different dynamical picture, was earlier put forward in
Ref.\cite{khasat}.

There is no evidence for a nuclear target suppression of inelastic $\jpsi$
{\em photo-} and {\em lepto}production. On the contrary, there is an
indication \cite{nmc} of a slight nuclear {\em enhancement} at $x_F
\simeq 0.7$ $(\alpha = 1.05 \pm 0.03)$, in stark contrast to the strong
nuclear suppression seen at this $x_F$ in hadroproduction
\cite{lourenco}. Due to the absence of comoving gluons and the validity
of the CSM we expect $\alpha \simeq 1$. The enhancement may signal a
slight antishadowing of the nuclear gluon distribution \cite{goupir}.

The COM and CEM assume that the process which turns the color octet
$\qpair$ into physical quarkonium is independent of the nature of the beam
and target. Both models expect photo- and hadroproduction of charmonium
to have the same target $A$-dependence, which should moreover equal that
of open charm.

\subsection{Dependence on $m$, $E_{CM}$ and $\ptr$} \label{sec23}

The available data shows that the `anomalies' of quarkonium production are
rather insensitive to the quark mass $m= m_c,\ m_b$ and to variations in
kinematic variables such as the total energy $E_{CM}$ and the quarkonium
transverse momentum $\ptr$.

The measured cross section of $\Upsilon(3S)$, which presumably is
directly produced, exceeds the CSM prediction by an order of magnitude
\cite{rev3,abe}, in analogy to the original `$\psi'$ anomaly'. The
$\Upsilon(1S)$ and $\Upsilon(2S)$ cross sections are more compatible with
the CSM, which predicts them to originate almost exclusively from
$P$-wave decays. The situation is thus similar to that of the
charmonium system before the $P$-wave contributions were experimentally
separated and found not to account for the bulk of the $\jpsi$ cross
section. It would obviously be very important to measure the directly
produced fractions of the $\Upsilon$ states.

All $\Upsilon(nS)$ states have similar nuclear target $A$-depen\-dence,
with $\alpha = 0.962 \pm 0.014$ \cite{alde}. The nuclear suppression is
thus smaller than that for charmonium discussed in Section~\ref{sec222},
but still significant compared to the Drell-Yan case. In our approach,
the smaller suppression for bottomonium is related to the effective
distance $\sim 1/\mu$ between the comoving gluon field and the
heavy quarks, which decreases as the inverse of the quark mass.

The $\Upsilon(3S)$ total cross section anomaly has been observed at both
fixed target \cite{alde,benrot} and collider energies \cite{abe}, \ie, for
$4.2 \lsim E_{CM}/M_{\Upsilon} \lsim 190$. Similarly, the nuclear target
suppression of charmonium production seems to be independent of the
projectile type ($\pi$ or $p$) and energy (for 150 GeV $<E_{LAB}<$ 800
GeV) \cite{badier,lourenco}. The discrepancy between the CSM and data on
direct $\jpsi$ production is somewhat larger at high transverse momentum
$(p_{\perp} \gg m)$ than for the total cross section. The
relative contribution of $P$-wave decays to $\jpsi$ production is roughly
independent of $\ptr$ \cite{rev3}.

The above features suggest that the anomalies observed in quarkonium
production are `leading twist' in the quark mass $m$, in the total energy
$E_{CM}$ and in $\ptr$, in the sense that the effects do not vanish as
inverse powers of any of those variables.

In the COM, the color octet contributions which account for direct
$\jpsi$ and $\psi'$ production scale by a factor $v^2$ relative to the
contributions from $P$-wave decays, and thus are relatively less
important for bottomonia than for charmonia. Thus $P$-wave decay
contributions dominate $\Upsilon(nS)$ production in the COM. In
particular, $\Upsilon(3S)$ production can only be understood by assuming
\cite{benrot,cholei} that it results from the decay of an (as yet
undetected) higher lying $P$-wave state.

\section{A Hard Rescattering Scenario} \label{sec3}

In this Section we address how the features (A -- C) of quarkonium
production, mentioned in Section~\ref{sec1} and discussed in more detail
in Section~\ref{sec2}, can be understood in a QCD framework. We put
forward a scenario which is consistent with those features, and which
forms the basis for the explicit model studied in Section~\ref{sec4}.

\subsection{The Basic Heavy Quark Creation Process} \label{sec31}

The hadroproduction of heavy quarks proceeds mainly via the gluon fusion
subprocess\footnote{We neglect light quark fusion $q\bar q \to \qpair$,
which is unimportant for the total cross section. Similarly, higher order
`gluon fragmentation' diagrams \cite{bryu} are irrelevant for quarkonia
produced with $\ptr \lsim m$.} $gg \to \qpair$. In our approach (as well
as in that of COM and CEM, but not of CSM) color neutralization occurs at
a time scale which is large compared to the time scale $1/m$ of the gluon
fusion process. This implies that the heavy
quarks are produced (nearly) on their mass shell.
\begin{figure}[h]
\center\leavevmode
\epsfxsize=6.2cm
\epsfbox{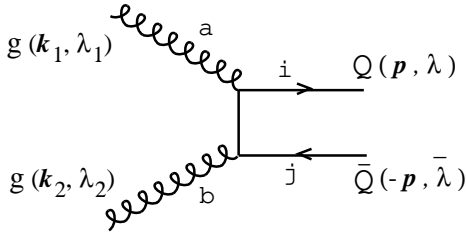}
\medskip
\caption{Notation for the CM amplitude $\Phi(\protect{\pvec})$. The spin
projections $\lambda_{1,2}= \pm 1$ and $\lambda,\bar\lambda =
\pm \slantfrac{1}{2}$ all refer to the $z$-axis, taken as the direction of
$\protect{\kvec_1}$. Only one of three contributing Feynman diagrams is shown.}
\end{figure}

A basic building block of our rescattering process of Fig.~2 is thus the
amplitude $\Phi^{[8]}$ of the $gg \to \qpair$ fusion process shown in
Fig.~3, evaluated at leading order in the heavy quark momentum $\pvec$
in the CM frame, and with the quark pair in a color octet state.

This amplitude is given in Table~\ref{tab1} in terms of the spin ($S$)
and angular momentum ($L$) of the $\qpair$ pair (see Section~\ref{sec4}, Eqs.
(\ref{spinampl}) and (\ref{partwave}) for definitions).

The gluon fusion amplitude $\Phi^{[1]}$ for $\qpair$ pairs in a color
singlet state, which is relevant for the CSM, can be obtained from
Table~\ref{tab1} by the substitutions $f_{abc} \to 0,\
d_{abc}T^c_{ij} \to \delta_{ab}\delta_{ij}/N_c$. It has two parts, with
spin and angular momentum $S=L=0$ and $S=L=1$, respectively. The
former can directly form $^1S_0$ ($\eta$) quarkonia, while the latter
couples to $^3P_{0,2}$ $(\chi_{0,2})$ states. The $\chi_1$ decouples
since  for $S=L=1$ the amplitude is symmetric in $L_z$ and $S_z$, whereas
the Clebsch-Gordan coefficients for the corresponding $(J=1, J_z)$  state
are antisymmetric in $L_z$ and $S_z$ (Yang's theorem).

The color octet amplitude $\Phi^{[8]}$ of Table~\ref{tab1} has
contributions from $S=L=0$, $S=L=1$ and $S=0, L=1$. The near-threshold
heavy quark cross section is dominantly $S=L=0$. In the CEM \cite{cem1}
the $S=0$ quark pair is assumed to turn into physical $S=1$ quarkonia
through soft, non-perturbative gluon interactions. This contradicts the
conservation of heavy quark spin in soft interactions, which is believed
to be a general feature of QCD and follows from the non-perturbative
concept of heavy quark symmetry \cite{isgurwise}. 
\begin{table}[htb]
\twocolumn[\hsize\textwidth\columnwidth\hsize\csname @twocolumnfalse\endcsname
\begin{tabular}{ccccc}
\multicolumn{2}{c}{{\LARGE\strut}\large{
$\left(\Phi^{[8]}\right)^{SS_z}_{LL_z}$}}
&\large{$S=0$} & \multicolumn{2}{c}{\large{ $S=1$}\ \ \ \ \ \ \ \ \ }\\
&&&\large{ $S_z = 0$} &\large{ $S_z = \pm 1$} \\
\hline
\multicolumn{2}{l}{{\LARGE\strut}\large{ $L=0$}} &\large{ $\lambda_1 d_{abc}
 \delta_{\lambda_1}^{-\lambda_2}$} &\large0 & \large0 \\
{\LARGE\strut}\large{ $L=1$} &
{\LARGE\strut}\large{ $L_z=0$} &\large{ $\frac{|\pvec|}{3m}\lambda_1
if_{abc} \delta_{\lambda_1}^{-\lambda_2}$} & \large{
$\frac{|\pvec|}{3m} d_{abc} \delta_{\lambda_1}^{-\lambda_2}$} & \large0 \\
& {\LARGE\strut}\large{ $L_z=\pm 1$} & \large0 & \large0 & \large{
$-\frac{|\pvec|}{3m} d_{abc} \left(\delta^{L_z}_{\lambda_1}
\delta^{S_z}_{\lambda_2} +
 \delta^{S_z}_{\lambda_1} \delta^{L_z}_{\lambda_2} \right)$} \\
\end{tabular}
\caption{The amplitude $\Phi^{[8]}$ of the $gg
\to Q\bar Q$ process to first order in $|\protect{\pvec}|/m$.}
\label{tab1}
\vskip2.0pc]
\end{table}
\eject
In the COM \cite{com}
the suppression of heavy quark spin-flip appears as extra
powers of the velocity $v \sim \as$ of the bound quarks. COM production
of $^3S_1$ states is suppressed by $v^4$ in the cross section, compared
to the  CSM.

\subsection{Scenario for Perturbative Rescattering} \label{sec32}

For the quarks to form a quarkonium bound state their relative momentum
must be of order $vm \ll m$. In current quarkonium production models
the quarks are created directly in the gluon fusion process with a
relative momentum of this magnitude. It is thus implicitly assumed that
all later interactions are soft, commensurate with the bound state
momentum scale.

Here we shall consider reinteractions with momentum transfers of
\order{\mu}, related to the hardness scale of the bremsstrahlung
field in the fusion subprocess. Since $v \ll \mu /m \ll 1$ the relative
momentum of the quarks created in the fusion process is fairly large. The
rescattering is hard and allows perturbative spin-flip interactions.

It might at first appear that the rescattering physics we have in mind is
contained in the loop correction (Fig.~1b) to the CSM lowest order process of
Fig.~1a. However, the logarithmic enhancements which favor collinear and
wee gluon bremsstrahlung are absent in the loop\footnote{Leading
logarithms in hard processes originate from tree diagrams. It is also
straightforward to verify the absence of logarithms directly from the
loop integral.}. Consequently, the loop momentum is of order $m$ and the
spatial size of the loop is of order $1/m$. The loop should be thought of
as a vertex correction to the primary process, not as the rescattering
envisioned in point (B) of Section~\ref{sec1}.

In order to have a rescattering which is well separated from the primary
fusion process we must assume that the initial state gluon radiation
gives rise to a (classical) color field $\Gamma$ which is comoving with
the $\qpair$ pair. The compact color octet $\qpair$ will dominantly
interact with $\Gamma$ as a color monopole (massive pointlike gluon). In
these relatively soft interactions the internal structure of the pair is
preserved, in particular it remains a color octet \cite{hope}.

The color structure and spin of the quark pair can, however, be changed in
a harder, color dipole interaction with the comoving field, as
depicted in Fig.~2. Here we consider only a single such interaction,
and evaluate it perturbatively. The main unknown in Fig.~2 is then the
postulated color field $\Gamma^{\mu}(\ell)$. Quarkonium production in
fact offers an opportunity of detecting whether such fields are created
in hard interactions.

The physical picture sketched above implies that the classical color
source $\Gamma^{\mu}(\ell)$ of gluons with momentum $\ell$ should
have the following properties.

\begin{itemize}
\item Since the field $\Gamma$ originates from bremstrahlung in the
gluon fusion subprocess it is independent of the beam and target.

\item
Since only those components of the radiated field which are comoving
with the quark pair are relevant, the spatial distribution of
$\Gamma$ is isotropic in the rest frame of the $\qpair$ pair.

\item The 3-momentum exchange $\lvec$ is of order $\mu$.
For the heavy quark propagators in Fig.~2 to be nearly on-shell
the energy component  satisfies $|\ell^0| \ll |\lvec|$
in the quarkonium rest frame. The field $\Gamma$ effectively acts as a
time-independent color source in the rescattering process.
\end{itemize}

By gauge invariance we have $\ell_\mu \Gamma^\mu =0$. Hence the field
$\Gamma^\mu$ can be expressed in terms of its transverse $(\lambda=\pm 1)$
and longitudinal $(\lambda=0)$ components as
\beq
\Gamma^\mu(\ell) = \sum_\lambda \varepsilon^\mu_{\lambda}(\ell)
\Gamma_\lambda(\ell)
\label{gammapol}
\eeq
where $\varepsilon^\mu_{\lambda}$ is the gluon polarization vector.
As we shall see in Section~\ref{sec4}, the quarkonium data requires that
$|\Gamma^0(\ell)| \ll |\bbox{\Gamma}(\ell)|$, which is equivalent to
$|\Gamma_{\lambda=0}(\ell)| \ll |\Gamma_{\lambda=\pm 1}(\ell)|$ since
$\ell_{\mu}\varepsilon^\mu_{\lambda} =0$ and
$|\ell^0| \ll |\lvec|$. Hence data suggests that the transverse components
$\Gamma_{\pm 1}$ of the color field dominate.

Quark pairs that in the gluon fusion process are created as color
singlets can evolve directly into $^1S_0$ and $^3P_{0,2}$ quarkonia,
as in the CSM. The influence of the color field $\Gamma$ on this process
is disruptive. To first order in the dipole interaction the pair turns
into a color octet, thus losing its overlap with the quarkonium wave
function.

Since the direct CSM process does not require any perturbative
rescattering, one might expect that it will dominate the contribution of
Fig.~2 for, \eg, $\chi_2$ production. However, direct
production of $P$-states is suppressed because the quarkonium wave
function vanishes at the origin. In the rescattering process the quark
pair is produced with a comparatively high relative momentum of
\order{\mu} and the spatial size of its wave function increases
before the rescattering.  The expansion factor $\rho$ depends on the
time interval between the gluon fusion and rescattering processes.
Our model can explain the relative rates of $\chi_{c2},\ \chi_{c1}$ and
$\jpsi$ production provided $\rho \simeq 3$. The expansion will
have a smaller effect on $S$-wave cross sections as long as the pair stays
compact compared to the size of the quarkonium wave function.

In the next Section we construct a simple, specific model for our scenario
and derive its quantitative predictions. We show how the observed fact
that the $\jpsi$ and $\psi'$ are produced unpolarized requires the field
$\Gamma$ to be dominantly transverse. We then have two free parameters,
the strength of the transverse field $\Gamma$ and the spatial expansion
parameter $\rho$, which as mentioned above is related to the time
interval between the quark creation and rescattering processes.

\section{Quarkonium Cross Sections and Polarization} \label{sec4}

\subsection{The Quarkonium Production Amplitude}  \label{sec41}

As illustrated in Fig.~2, the production amplitude can be viewed as a
transition from a color octet $\qpair$ state, created in the gluon fusion
process and described by the wave function $\Phi$, to a quarkonium
state specified by the wave function $\Psi$. The transition is mediated
by a gluon exchange $\R$ with the color field $\Gamma$. The
transition amplitude can be expressed as an overlap integral,
\beqa
\M&=& \sum_{L_z,S_z} \langle LL_z; SS_z |JJ_z \rangle
\sum_{\llb,\ssb} \int \frac{d^3\pvec}{(2\pi)^3}
\frac{d^3\qvec}{(2\pi)^3} \nonumber \\ &\times&\Phi_\llb^{[8]}(\pvec)\
\R_{\llb,\ssb}(\pvec,\qvec)\ {\Psi_\ssb^{L_zS_z}(\qvec)}^*
\label{prodampl}
\eeqa
which can readily be derived starting from the usual Feynman
formulation. We work in a non-relativistic approximation where all quark
lines are on-shell and have 3-momenta much smaller than the quark mass
$m$. The $z$-axis is taken along the direction of the initial gluon
momentum $\kvec_1$ in Fig.~2. The orbital angular momentum and spin
components of the quarkonium state are denoted by $L_z$ and
$S_z$, whereas $\lambda, \bar\lambda$ and $\sigma,\bar\sigma$ are the spin
projections of the $Q$ and $\bar Q$ along the $z$-axis before and after
the rescattering, respectively. The relative momentum between the $Q$ and
$\bar Q$ is $2\pvec$ before and $2\qvec$ after the rescattering, while
$\lvec = \bbox{P_f}-\bbox{P_i}$ is the momentum of the exchanged gluon. In
the quarkonium rest frame $\bbox{P_f}=0$ and $\lvec=-\bbox{P_i}$.

\subsubsection{The gluon fusion amplitude $\Phi$}  \label{sec411}

A standard calculation of the gluon fusion process $gg \to\qpair$ of
Fig.~3 gives the $\qpair$ wave function, to first order in the quark CM
momentum $\pvec$, as
\begin{mathletters} \label{phiampl}
\beqa
\Phi_{\llb}^{[8]}(\pvec) &=& g^2 T^c_{ij} \left\{ i \left(d_{abc}+
\frac{p^z}{m}if_{abc}\right) (\epsvec_1 \times \epsvec_2)^z
\delta_{\lambda}^{-\bar\lambda}\right. \nonumber\\
&+&\left. 2 \lambda d_{abc}\frac{1}{m} \left[ \epsvec_1 \cdot
\epsvec_2\,\, \pvec \cdot \evec(0)^* \delta_{\lambda}^{-\bar\lambda}
\right.\right. \label{phioctet}\\
&-&\left.\left.\left(\epsvec_1 \cdot \pvec\,\,\epsvec_2 \cdot
\evec(2\lambda)^* +\epsvec_2 \cdot \pvec\,\, \epsvec_1 \cdot
\evec(2\lambda)^*\right)
\sqrt{2}\delta_{\lambda}^{\bar\lambda} \right] \right\}
\nonumber
\eeqa
\beq
\Phi_{\llb}^{[1]} = \Phi_{\llb}^{[8]} \left\{f_{abc}\to 0,\
d_{abc}T^c_{ij} \to \delta_{ab}\delta_{ij}/N_c \right\}
\label{phisinglet}
\eeq
\end{mathletters}
As indicated, the amplitude $\Phi^{[1]}$ for a color singlet pair can be
obtained from the color octet amplitude $\Phi^{[8]}$ by a trivial
substitution. The incoming gluons have colors $a,b$, momenta
$m(1,0_\perp,\pm 1)$ and polarization vectors $\epsvec_1,\epsvec_2$. The
quarks $Q,\bar Q$ of colors $i,j$ have momenta $(m,\pm\pvec)$ and spin
projections $\lambda, \bar\lambda$ along the $z$-axis. The spin
polarization vector $\evec$ is defined conventionally by
\begin{mathletters} \label{polvec}
\beqa
e(\pm 1) &=& (0,\mp 1,-i,0)/\sqrt{2} \label{etrans}\\
e(0) &=& (0,0,0,1) \label{elong}
\eeqa
\end{mathletters}

Lorentz transforming from the $gg \to \qpair$ CM to the quarkonium rest
frame shifts any four-vector by an amount of \order{|\lvec|/m}. In
particular, $(\epsvec_1 \times
\epsvec_2)^z$ is shifted to $(\epsvec_1 \times \epsvec_2)^z
\left[1+\morder{\lvec^2/m^2}\right]$. To the accuracy of our calculation
we can ignore the boost and use the CM amplitude (\ref{phioctet})
directly in \eq{prodampl}.

It is instructive to express the amplitudes (\ref{phiampl}) in terms of
the spin $S$ and orbital angular momentum $L$ of the $\qpair$ pair. The
amplitudes $\Phi^{SS_z}$ of definite $\qpair$ spin are defined by
\begin{mathletters} \label{spinampl}
\beqa
\Phi^{00} &=& -\left( \Phi_{\half,-\half} +
\Phi_{-\half,\half} \right)/\sqrt{2}  \label{spinsinglet} \\
\Phi^{10} &=& -\left( \Phi_{\half,-\half} -
\Phi_{-\half,\half} \right)/\sqrt{2}  \label{triplet0} \\
\Phi^{1,\pm1} &=& \pm \ \Phi_{\pm\half,\pm\half}  \label{triplet1}
\eeqa
\end{mathletters}

The angular momentum components are defined via a partial wave
expansion,
\beq
\Phi^{SS_z} = \sum_{LL_z} \sqrt{4\pi(2L+1)} Y_L^{L_z}(\theta,\phi)
\sqrt{2}g^2 T^c_{ij}\Phi^{SS_z}_{LL_z}  \label{partwave}
\eeq
where $Y_L^{L_z}$ are standard spherical harmonics \cite{pdg}. The
amplitudes $\Phi^{SS_z}_{LL_z}$ are given in Table~\ref{tab1} (see
also \cite{cholei}).

\subsubsection{The rescattering kernel $\R$}  \label{sec412}

The rescattering kernel $\R$ has two terms, describing gluon
scattering from the quark and from the antiquark,
\beqa
\R_{\llb,\ssb}(\pvec,\qvec) &=& (2\pi)^3 \delta^3(\pvec-\qvec
+\frac{\lvec}{2}) \R^Q_{\llb,\ssb}(\lvec,\qvec)  \nonumber \\
&+& (2\pi)^3 \delta^3(\pvec-\qvec
-\frac{\lvec}{2}) \R^{\bar Q}_{\llb,\ssb}(\lvec,\qvec)
\label{rkernel}
\eeqa
where
\beqa
\R^Q_{\llb,\ssb}(\lvec,\qvec) &=& -ig \frac{\Gamma_\mu(\lvec)}{\lvec^2}
\delta_{\bar\lambda}^{\bar\sigma} \frac{1}{2m} \bar u(\qvec,\sigma)
\gamma^\mu u(\qvec-\lvec,\lambda) \nonumber \\
\R^{\bar Q}_{\llb,\ssb}(\lvec,\qvec) &=& +ig
\frac{\Gamma_\mu(\lvec)}{\lvec^2}
\delta_{\lambda}^{\sigma} \frac{1}{2m} \bar v(-\qvec-\lvec,\bar\lambda)
\gamma^\mu v(-\qvec,\bar\sigma)  \nonumber \\  \label{rexpr}
\eeqa
To first order in $\lvec/m,\ \qvec/m$ this reduces to
\beqa
&&\R^Q_{\llb,\ssb}(\lvec,\qvec) = ig \frac{\Gamma_\mu(\lvec)}{\lvec^2}
\delta_{\bar\lambda}^{\bar\sigma} \nonumber \\
&&\times \left\{-\delta_0^\mu \delta_\lambda^\sigma + \delta_i^\mu
\frac{1}{2m} \left[\delta_\lambda^\sigma (\ell-2q)^i + i\epsilon_{ijk}
\ell^j \chi_\sigma^\dagger \sigma_k \chi_\lambda \right] \right\}
\nonumber \\
&&\R^{\bar Q}_{\llb,\ssb}(\lvec,\qvec) =
ig\frac{\Gamma_\mu(\lvec)}{\lvec^2} \delta_{\lambda}^{\sigma} \nonumber\\
&&\times \left\{\delta_0^\mu \delta_{\bar\lambda}^{\bar\sigma} +
\delta_i^\mu \frac{1}{2m} \left[\delta_{\bar\lambda}^{\bar\sigma}
(-\ell-2q)^i + i\epsilon_{ijk}
\ell^j \chi_{-\bar\lambda}^\dagger \sigma_k \chi_{-\bar\sigma}^{} \right]
\right\} \nonumber \\ \label{reval}
\eeqa
Here $\sigma_k$ are the Pauli matrices and $\chi_\lambda$
the spinors $\chi_+^\dagger = \left( 1\ 0 \right)$, $\chi_-^\dagger =
\left( 0\ 1 \right)$.

We note that the rescattering kernel $\R$ consists of two parts. A
$\mu=0$ spin-conserving part which is of \order{1}, and a $\mu=i$ part
which also contains spin-flip and is of \order{\lvec/m,\,\qvec/m}.

\subsubsection{The quarkonium wave function $\Psi$}  \label{sec413}

The quarkonium wave function $\Psi$ in \eq{prodampl} is
\beq
{\Psi_\ssb^{L_zS_z}(\qvec)}^* = \Psi_{LL_z}^* (\qvec) \frac{1}{2m} \bar
v(-\qvec,\bar\sigma) \bar{\Ps}_{SS_z}(\qvec,-\qvec) u(\qvec,\sigma)
\label{psiampl}
\eeq
where $\Psi_{LL_z} (\qvec)$ is the usual non-relativistic bound state
wave function and the spin-projection operator $\bar{\Ps}_{SS_z}$ is given
by \cite{spinproj}
\beqa
\bar{\Ps}_{SS_z}(\pvec_1,\pvec_2) &=& \frac{1}{2m}
\sum_{\lambda_1,\lambda_2}
\langle \shalf\lambda_1,\shalf\lambda_2 | SS_z \rangle
v(\pvec_2,\lambda_2)
\bar u(\pvec_1,\lambda_1) \nonumber \\
&=& \frac{1}{2m} \frac{\Sslash{p}_2 - m}{\sqrt{p_2^0+m}}\bar{\Pi}_{SS_z}
\frac{\gamma^0+1}{2\sqrt{2}}  \frac{\Sslash{p}_1 + m}{\sqrt{p_1^0+m}}
\nonumber \\
&\simeq& \frac{1}{(2m)^2\sqrt{2}} (\Sslash{p}_2 - m) \bar\Pi_{SS_z}
(\Sslash{p}_1 + m) \label{projoper} \\
\bar\Pi_{SS_z} &=& \left\{
\begin{array}{cc}
-\gamma_5 & (S=0) \\
\Sslash{e}(S_z)^* & (S=1) \\
\end{array} \right. \label{gamproj}
\eeqa
Here we used $p_1 = P_f/2+q = (m,\qvec)$, $p_2 = P_f/2-q = (m,-\qvec)$
and terms of order $q^2$ were neglected in the last line of
\eq{projoper}. A simple calculation gives for the $S=1$ case
\beq
{\Psi_\ssb^{L_zS_z}(\qvec)}^* = \Psi_{LL_z}^* (\qvec) \frac{1}{\sqrt{2}}
\evec(S_z)^* \cdot \chi_{-\bar\sigma}^\dagger \bbox{\sigma} \chi_\sigma
\label{spinoneampl}
\eeq

It is now straightforward to use Eqs.~(\ref{phiampl}),
(\ref{reval}) and (\ref{spinoneampl}) in \eq{prodampl}. We contract over
both pairs of indices $\lambda,\bar\lambda$ and $\sigma, \bar\sigma$,
and work to first order in $\lvec/m,\ \qvec/m$. The color structure
of the field $\Gamma$ in \eq{reval} is made explicit by
\beq
\Gamma_\mu(\lvec) \to \frac{1}{\sqrt{3}} T_{j'i'}^d \Gamma_\mu^d(\lvec)
\label{gamcolor}
\eeq
where $d$ is the color index of the rescattering gluon and $i',j'$ are
the colors of the quark and antiquark just before the rescattering. The
factor $1/\sqrt{3}$ is from the color singlet bound state wave function.

Finally, we use for $S$-wave states
\beqa
\int \frac{d^3\qvec}{(2\pi)^3} \Psi_{00}^*(\qvec) &=&
\frac{R_0}{\sqrt{4\pi m}}  \nonumber  \\
\int \frac{d^3\qvec}{(2\pi)^3}\, \qvec \Psi_{00}^*(\qvec) &=& \bbox{0}
\label{swavenorm}
\eeqa
where $R_0$ is the value of the $S$-wave function at the origin. For
$P$-wave states
\beqa
\int \frac{d^3\qvec}{(2\pi)^3} \Psi_{1L_z}^*(\qvec) &=& 0  \nonumber \\
\int \frac{d^3\qvec}{(2\pi)^3}\, \qvec \Psi_{1L_z}^*(\qvec) &=&
i\sqrt{\frac{3}{4\pi m}}{R_1}' \evec(L_z)^*
\label{pwavenorm}
\eeqa
where ${R_1}'$ is the derivative of the $P$-wave function at the origin.

\subsection{The $^3S_1$ quarkonium cross section} \label{sec42}

We find for the $^3S_1$ quarkonium production amplitude
\beqa
\M(&^3S_1&,S_z) = D^d \frac{2ig^3R_0}{\sqrt{6\pi m^3}}\frac{1}{\lvec^2}
\left\{ i\lambda_1 \delta_{\lambda_1}^{-\lambda_2} \bbox{\Gamma}^d
(\lvec) \times \lvec \cdot \evec(S_z)^* \right. \nonumber \\
&+& \left. \Gamma_0^d(\lvec) \left[ -\delta_{\lambda_1}^{-\lambda_2}
\ell^z \delta_{S_z}^0 + \evec(\lambda_1) \cdot \lvec
\delta_{S_z}^{\lambda_2} + \evec(\lambda_2) \cdot \lvec
\delta_{S_z}^{\lambda_1} \right] \right\} \nonumber \\
\label{swaveampl}
\eeqa
where the color factor is proportional to $d_{abc}$,
\beq
D^d = \half d_{abc} T_{ij}^c \, T_{j'i'}^d  \label{scolorfact}
\eeq
and $\lambda_1,\lambda_2$ are the $z$-components of the spins of the
incoming gluons $(\lambda_{1,2} = \pm 1)$.
As can be readily inferred from the form of Eqs.~(\ref{phiampl}) and
(\ref{reval}) the 1st and 2nd term of \eq{swaveampl} correspond,
respectively, to
\begin{itemize}
\item the production of the $\qpair$ pair in an $S=L=0$ state,
followed by a spin-flip interaction with a gluon from the color field
$\bbox{\Gamma}$, and
\item the production of the pair in an $S=L=1$ state, followed by a
spin-conserving interaction with a gluon from the color field
$\Gamma^0$.
\end{itemize}

The result for the amplitude squared, summed over the quark color
indices $i,j$ and averaged over $i',j'$ and the gluon spin components
$\lambda_1,\lambda_2 = \pm 1$, is for a given spin component $S_z$ of
the quarkonium
\beqa
&\sum& \overline{|\M(^3S_1,S_z)|^2} =
\frac{2}{3}\frac{N^2-4}{2N^3} \frac{g^6 R_0^2}{6\pi m^3} \nonumber \\
&\ & \times {\frac{1}{\lvec^4}} \,\left\{ {\left| \bbox{\Gamma}^d(\lvec)
\times \lvec \right|}^2
+ c(S_z) \lvec^2 {\left| \Gamma_0^d(\lvec) \right|}^2 \right\}
\label{samplsq}
\eeqa
where $c(S_z=0) = 1$ and $c(S_z= \pm 1) = 3$. We have used $D^d D^{d'}
= \delta_d^{d'} (N^2-4)/2N$ and the fact that the color field $\Gamma$
is isotropically distributed. Thus, for instance,
\beqa
{\left( \bbox{\Gamma}^d(\lvec) \times \lvec \right)}^x \,
{\left( \bbox{\Gamma}^d(\lvec) \times \lvec \right)}^y \to 0
\nonumber \\
{\left| {\left( \bbox{\Gamma}^d(\lvec) \times \lvec \right)}^z
\right|}^2 \to \frac{1}{3} {\left| \bbox{\Gamma}^d(\lvec) \times \lvec
\right|}^2  \nonumber \\
\eeqa

The fixed target data \cite{badier,poln,heinrich} shows that the
polarization of both the $\jpsi$ and the $\psi'$ is small and consistent
with zero. We see from \eq{samplsq} that the condition for the (directly
produced) $^3S_1$ quarkonium states to be unpolarized is
\beq
\left| \Gamma_0^d(\lvec) \right| \ll
\left| \bbox{\Gamma}^d(\lvec) \times \hat{\lvec} \right|
\label{transgam}
\eeq
In the following we shall therefore take
\beq
\Gamma_0^d(\lvec) = 0  \label{longgam}
\eeq
which implies (\cf\ \eq{gammapol}) that the color field $\Gamma$ is
made of transversely polarized gluons (in the $\qpair$ rest frame), and
that $\qpair$ pairs that form $^3S_1$ quarkonia are created with
$S=L=0$.

We thus have, for $^3S_1$ production via gluon fusion,
\beq
\sum \overline{|\M(^3S_1,S_z)|^2} =
\frac{160}{243} \pi^2 \as^3 \frac{R_0^2}{m^3} \frac{{\left|
\bbox{\Gamma}^d(\lvec) \times \lvec \right|}^2}{\lvec^4}
\label{scross}
\eeq
The cross section will involve an integral of this expression over
$\lvec$. Hence the weighted average
\beq
\overline{\bbox{\Gamma}^2} \equiv
\int \frac{d^3\lvec}{(2\pi)^3} \frac{{\left|
\bbox{\Gamma}^d(\lvec) \times \lvec \right|}^2}{\lvec^4}
\label{avegam}
\eeq
is the main free parameter of our model, which can be determined, \eg,
using the measured (direct) $\jpsi$ cross section. The $\psi'$ cross
section then satisfies \eq{psiratio}, as discussed in
Section~\ref{sec2}. As we shall see below, the $P$-wave production
cross sections are also proportional to $\overline{\bbox{\Gamma}^2}$,
but in that case an additional parameter enters, which is related to the
length of the time interval between the gluon fusion and rescattering
processes.

Since the color field $\Gamma$ arises through gluon radiation in the
primary gluon fusion process it should be the same for all charmonium
(as well as open charm) amplitudes. In the production of $b\bar b$
pairs, on the other hand, the gluon radiation and thus also
the color field parameter (\ref{avegam}) will be different.

\subsection{The $P$-wave quarkonium cross sections}  \label{sec43}

The $P$-wave production amplitudes are obtained similarly to the
$S$-wave ones, by substituting Eqs.~(\ref{phiampl}),
(\ref{reval}) and (\ref{spinoneampl}) into \eq{prodampl}. However,
since the $P$-wave function vanishes at the origin it is now important
to take into account the spatial expansion of the $\qpair$ pair
between its creation in the gluon fusion process and its reinteraction
with the color field $\Gamma$.

The production amplitude (\ref{prodampl}) can equivalently be expressed
in coordinate space as an overlap between the wave function of the
quark pair $(\Phi)$ and that of the quarkonium bound state $(\Psi)$ at
the {\em same} spatial separation,
\beqa
\M &=& \int d^3 \xvec_1 \, d^3 \xvec_2 \, d^3 \xvec_3 \ \Phi(\xvec_1 -
\xvec_2)\, \Gamma(\xvec_3)\, \Psi^*(\xvec_1 - \xvec_2) \nonumber \\
&\times& \left[D(\xvec_1 - \xvec_3)+ D(\xvec_2 - \xvec_3) \right]
 \nonumber \\ &\times&
\exp\left[-i(\Pvec_f-\Pvec_i) \cdot \frac{\xvec_1 + \xvec_2}{2}\right]
\label{coordampl}
\eeqa
Here $\xvec_1, \xvec_2$ are the positions of the heavy quarks at the
reinteraction time, and $D$ stands for the exchanged gluon propagator.
Since $\Psi(\bbox{0})=0$ for $P$-wave quarkonia the amplitude
(\ref{coordampl}) is sensitive to the spatial extent of the quark wave
function $\Phi(\xvec_1 -\xvec_2)$ at the time of reinteraction.

According to our discussion in Section~\ref{sec32}, the $\qpair$ pair is
created in the gluon fusion process with a size $\sim 1/m$ and a
relative momentum of \order{\mu}, which is large compared to the
relative momentum $\qvec$ of the quarkonium bound state,
$\mu \sim |\lvec| \gg |\qvec|$. In the (proper) time interval
$\tau$ before the pair reinteracts with $\Gamma$ it thus expands a
distance
\beq
\Delta |\xvec| \simeq \frac{\mu}{m}\tau 
\label{expdist}
\eeq
The expansion time must be at least of the order of the spatial size of
the comoving field, $\tau \gsim 1/\mu$, hence $\Delta |\xvec|
\gsim 1/m$ is comparable to (or even larger than) the initial size
of the pair. We shall parametrize the expansion by a factor $\rho> 1$,
and rescale the initial coordinate space gluon fusion amplitude
accordingly,
\beq
\Phi(\xvec_1 -\xvec_2) \to \frac{1}{\rho^3}
\Phi\left(\frac{1}{\rho}(\xvec_1 -\xvec_2)\right)
\label{phirescale}
\eeq
where the factor $1/\rho^3$ preserves the normalization of the squared
wave function. In momentum space, the rescaling (\ref{phirescale})
implies
\beq
\Phi_{\llb}(\pvec) \to \Phi_{\llb}(\rho\pvec)  \label{momrescale}
\eeq
in \eq{phiampl}.

Intuitively it is clear that the rescaling of the argument of $\Phi$ by
$\rho$ will increase the overlap with the $P$-wave quarkonium wave
function by the same factor (and hence result in a $\rho^2$ enhancement
of the cross section). Conversely, the effect for $S$-waves vanishes in
the limit where the quarkonium bound state radius $\gg \rho/m$.
Formally, the enhancement of the $P$-wave amplitude can be seen from
\eq{phioctet}, where the $f_{abc}$ part (which due to charge
conjugation invariance contributes to $P$-wave production) is {\em
linear} in $\pvec$, and thus gets multiplied by $\rho$ in the rescaling
(\ref{momrescale}).

We find for the spin triplet $P$-wave production amplitude, taking into
account \eq{longgam}\footnote{When \eq{longgam} is satisfied, the
rescattering kernel $\R$ is linear in the small quantities
$\lvec/m,\,\qvec/m$ (see \eq{reval} with the $\mu=0$ term removed).
For $^3P_J$ production, the only contributing part in $\Phi^{[8]}$
of \eq{phioctet} is the term $\propto f_{abc}$, which is also linear.
The amplitude (\ref{pampl}) thus arises from a {\it quadratic} term
$\sim \qvec\cdot\lvec/m^2$, however our {\it linear} approximation in
$\lvec/m,\,\qvec/m$ is perfectly justified.},
\beqa
&\M&(^3P_J,J_z) = \rho F^d \frac{-2ig^3\sqrt{3}R_1'/m}{\sqrt{6\pi m^3}}
\frac{1}{\lvec^2} i\lambda_1 \delta_{\lambda_1}^{-\lambda_2}
 \nonumber\\
& \times& \left[ \bbox{\Gamma}^d(\lvec) \times \lvec \right]^i
\sum_{L_zS_z} \langle LL_z; SS_z |JJ_z \rangle e^z(L_z)^* e^i(S_z)^*
\nonumber \\ \label{pampl}
\eeqa
where
\beq
F^d = \half f_{abc} T_{ij}^c\, T_{j'i'}^d   \label{ffactor}
\eeq
Here the $\qpair$ pair is created with $S=0,\ L=1$ and experiences a
spin-flip interaction with the color field $\Gamma$. An explicit
expression for the spin sum in the last factor of \eq{pampl} may be
found in Ref.~\cite{spinproj}. Thus,
\beqa
\M(&^3P_J&,J_z) = \rho F^d \frac{2ig^3R_1'/m}{\sqrt{6\pi
m^3}}\frac{1}{\lvec^2} i\lambda_1 \delta_{\lambda_1}^{-\lambda_2}
\nonumber \\ &\times& \left\{
\begin{array}{cc}
\left[ \bbox{\Gamma}^d(\lvec) \times \lvec \right]^z & (J=0) \\
-i\sqrt{\frac{3}{2}} \left[ \evec^*(J_z) \times \left(
\bbox{\Gamma}^d(\lvec)\times \lvec \right) \right]^z & (J=1) \\
-\sqrt{3} e_{3i}^*(J_z) \left[ \bbox{\Gamma}^d(\lvec) \times \lvec
\right]^i & (J=2)
\end{array} \right.
\label{3pjampl}
\eeqa
The polarization tensor $e_{\mu\nu}$ for a $J=2$ system is given in
Ref.~\cite{cholei}.

The $P$-wave amplitude $\M(^3P_J,J_z)$ depends on the (isotropic) color
field $\Gamma$ through the vector $\bbox{\Gamma}^d(\lvec) \times \lvec$.
In the amplitude squared, averaged over the incoming gluon spins and
over the momentum transfer $\lvec$, the $\Gamma$-dependence thus enters
through the same parameter as for $S$-waves,
$\overline{\bbox{\Gamma}^2}$ of \eq{avegam}. Hence the relative
production rates and polarizations of all $^3P_J$ quarkonia are
predicted, and their cross sections can be compared to those of the
$^3S_1$ states in terms of the expansion parameter $\rho$. The ratios
$\sigma(^3P_J,J_z)/\sigma_{dir}(^3S_1)$ are given in Table~\ref{tab2},
where $\sigma_{dir}(^3S_1)$ is the total (unpolarized) $^3S_1$ cross
section calculated in Section~\ref{sec42} (\cf\ \eq{scross}). All cross
sections satisfy $\sigma(^3P_J,J_z) = \sigma(^3P_J,-J_z)$. The
effective parameter $r$ of Table~\ref{tab2} is defined as
\beq
r= \frac{3}{5} \rho^2 \left(\frac{R_1'/m}{R_0}\right)^2
\simeq \left\{
\begin{array}{dcc}
2.5 \, 10^{-2}\, \rho^2 & \hspace{.3cm} & (c\bar c) \\
6.5 \, 10^{-3}\, \rho^2 & \hspace{.3cm} & (b\bar b) \\
\end{array} \right.
\label{rparam}
\eeq
where the numerical values are from Ref.~\cite{eiqu}, with $m_c =
1.5$ GeV and $m_b = 4.5$ GeV. Note that Table~\ref{tab2} refers only to
quarkonium hadroproduction through our rescattering process (\cf\ Fig.
2). The CSM mechanism may contribute significantly to
$^3P_2$ production (see below). Table~\ref{tab2} is not
relevant for quarkonium photoproduction, to which our process does not
contribute since $\Gamma=0$ (in the photon fragmentation region) due
the absence of gluon radiation from the beam photon.

\begin{table}[htb]
\begin{tabular}{ccccccccc}
{\LARGE\strut} & {\large $^3P_0$} & \hspace{.2cm} &
\multicolumn{2}{c}{{\large $^3P_1$}} &
\hspace{.2cm} & \multicolumn{3}{c}{{\large $^3P_2$}} \\
{\LARGE\strut} {\large $J_z$} & {\large 0} && {\large 1} &{\large 0} &&
{\large 2} & {\large 1} & {\large 0} \\
\hline
{\LARGE\strut} {\large $\frac{\sigma(^3P_J,J_z)}
{\sigma_{dir}(^3S_1)}$} & {\large $r$} && {\large $\frac{3}{2}r$} &
{\large 0} && {\large 0}  & {\large
$\frac{3}{2}r$} & {\large $2r$} \\
{\LARGE\strut} {\large $\lambda_{indir}(^3S_1)$} & {\large 0} &&
{\large $-\frac{1}{3}$} & {\large 1} && {\large 1} &
{\large $-\frac{1}{3}$} & {\large $-\frac{3}{5}$} \\
\end{tabular}
\caption{Relative $^3P_J$ cross sections and induced $J/\psi$
polarizations.}
\label{tab2}
\end{table}

As can be seen from Table~\ref{tab2}, the total $P$-wave rates satisfy
\beq
\sigma(\chi_0)\,:\,\sigma(\chi_1)\,:\,\sigma(\chi_2) = 1:3:5
\label{chirates}
\eeq
There is no data on $\sigma(\chi_{c0})$, but our $\chi_{c1}/\chi_{c2}$
ratio is consistent with the value $0.6 \pm 0.3$ measured in $\pi
N$ collisions \cite{e705,e672}. The experimental ratio allows a CSM
contribution to $\chi_{c2}$ production which is about equal to the one
given in Table~\ref{tab2} ($\sigma(\chi_{c1})$ is small in the CSM).
There is no experimental information on the polarization of the $^3P_J$
states. Nevertheless, it is interesting to note that we find
$\sigma(^3P_2, J_z= \pm 2)= 0$, contrary to the CSM where {\em only} this
polarization is produced \cite{vhbt}.

The measured \cite{e705} cross section ratio\footnote{Ref.~\cite{e672}
gives a larger value $3.4\pm 0.9 \pm 0.5$ for this ratio.}
\beq
\frac{\sigma(\chi_{c2})}{\sigma_{dir}(\jpsi)} = 5 r \simeq 1.8 \pm .4
\eeq
implies using \eq{rparam} a value $\rho=2.7 \ldots 3.8$, where the
lower value corresponds to the CSM contributing 50\% of the
$\chi_{c2}$ rate. The rather large value $\rho \sim 3$ of the expansion
parameter is a consequence of the theoretical suppression of $^3P_J$
production due to the vanishing of the $P$-wave function at the origin
(\cf\ \eq{rparam}) and the fact that the measured $\sigma(\chi_{c1,2})$
nevertheless are similar to $\sigma_{dir}(\jpsi)$.

The radiative decays $\chi_{c1,2} \to \jpsi + \gamma$ contribute
\cite{e705,e672} $\sim 40\%$ of the total $\jpsi$ cross section. Since the
$\chi_c$ states are polarized the indirectly produced $\jpsi$'s will in
general be polarized as well. The polarization is conventionally
parametrized in terms of a parameter $\lambda$ in the $\jpsi \to
\mu^{+}\mu^{-}$
decay angular distribution (we use the Gottfried-Jackson frame, where
the $z$-axis in the $\jpsi$ rest frame is taken along the beam
direction),
\beqa
\frac{d\sigma}{d\cos\theta_\mu} &\propto& 1+\lambda \cos^2\theta
\nonumber \\
\lambda &=& \frac{\sigma(S_z=+1)-\sigma(S_z=0)}
{\sigma(S_z=+1)+\sigma(S_z=0)}  \label{lamdef}
\eeqa
As discussed above, the condition (\ref{longgam}) implies that the
directly produced $\jpsi$'s are unpolarized, $\lambda_{dir}=0$.
A radiative decay contributes to the indirect $^3S_1$ cross section
according to \cite{tava}
\beqa
\sigma_{indir}(&^3S_1&,S_z) = \text{Br}(^3P_J \to ^3S_1 + \gamma)
\nonumber \\
&\times& \sum_{J_z} \left| \langle JJ_z | 1(J_z-S_z); 1S_z \rangle
\right|^2 \sigma(^3P_J,J_z) \nonumber \\
\label{indcross}
\eeqa

We show in Table~\ref{tab2} the $\jpsi$ polarization parameter
$\lambda_{indir}(^3S_1)$ induced by the radiative decay of a
$^3P_J$ state of given $|J_z|$. It happens that the induced $\jpsi$
polarization is longitudinal $(\lambda < 0)$ for all the $^3P_J$ states
which are produced by our mechanism. Using branching fractions
$\text{Br}(\chi_{cJ} \to \jpsi + \gamma)$ of 27.3\% and 13.5\% for the
$\chi_{c1}$ and $\chi_{c2}$, respectively \cite{pdg}, we estimate an
overall polarization parameter\footnote{The inclusion of the unpolarized
$\psi' \to \jpsi$ contribution does not change the numerical value of
$\lambda$ significantly.}
$\lambda(\jpsi) \simeq -0.14$.
The measurements \cite{badier,poln} tend to prefer a value closer
to zero. There is thus room for some $\chi_{c2}$ production via the CSM
mechanism, which gives rise to fully transverse $(\lambda=1)$ $\jpsi$'s.
For $\sigma_{CSM}(\chi_{c2}) \simeq \half \sigma_{tot}(\chi_{c2})$ the
total $\lambda(\jpsi) \simeq -0.02$.

The $\psi'$ is only produced directly and in our mechanism is
unpolarized for a color field $\Gamma$ satisfying \eq{longgam}. This is
consistent with the experimental value \cite{heinrich} $\lambda(\psi') =
0.02 \pm 0.14$.

\section{Conclusions} \label{sec5}

Our motivation for investigating the reinteraction scenario of Fig.~2
for quarkonium hadroproduction was due both to regularities in the data
and to shortcomings of alternative  mechanisms, as explained in
Section~\ref{sec2}. Here we shall briefly comment on some aspects of our
results.

We made several simplifying assumptions, some of which may need to be
modified in future studies. In particular,
\begin{enumerate}
\item We considered only a single hard reinteraction with the comoving
field $\Gamma$.
\item We assumed $\Gamma$ to be isotropic and independent of time.
\item We assumed that only the origin of the quarkonium wave function
is relevant in the overlap integral (\ref{coordampl}).
\end{enumerate}
These assumptions seem reasonable in a first attempt to consider the
effects of reinteractions. We believe that further systematic studies of
the environment of partons created in hard collisions are called for.
Besides quarkonium production, the flavor and azimuthal angle
correlations as well as the spin dependence of open heavy flavor
production should be informative.

Our scenario for quarkonium production was based on several
striking features of the data, which we interpreted as the properties
A -- C listed in Section~\ref{sec11}. We also built on the
extensive experience gained from previous model studies of quarkonium
production. Many of the `successes' of the present approach were
thus built in from the start. Nevertheless, we find it non-trivial and
interesting that so many observed features can be qualitatively understood
in a simple theoretical framework. We also found more detailed
consequences of our model which could not be anticipated. Let us
mention two successes and one difficulty:

\begin{itemize}
\item {\em The $\sigma(\chi_{c1})/\sigma(\chi_{c2})$ ratio is consistent
with data.} This ratio is found to be much too low, compared to data, in
both the color singlet (CSM) \cite{vhbt} and color octet (COM)
\cite{benrot} approaches.
\item {\em The (non-)polarization of the $\jpsi$ and $\psi'$.} We find,
in agreement with data, that the directly produced $^3S_1$ states are
unpolarized (at moderate $x_F$), provided that the reinteraction is
dominated by
transverse gluon exchange, \cf\
\eq{transgam}. In both the CSM and COM the $\jpsi$'s are produced with
transverse polarization \cite{vhbt,benrot,tava}. The color evaporation
model (CEM) postulates that soft interactions flip the heavy quark
spin, in violation of heavy quark symmetry \cite{isgurwise}.
\item {\em Spatial expansion of the heavy quark pair.} The
initially compact pair expands by a factor $\rho$ before its
reinteraction with the color field creates an overlap with the
quarkonium wave function. We need $\rho \simeq 3$ to fit the observed
relative rates of $\jpsi$ and $\chi_{c2}$. Such a large expansion of the
quark pair may be inconsistent with our approximation, which considers
only the quarkonium wave function at the origin.
\end{itemize}

\eject
\section*{Acknowledgments}
We are grateful for helpful discussions with, in particular,
S. J. Brodsky, D. Kharzeev, K. Redlich, H. Satz, R. Thews, R. Venugopalan,
M. V\"anttinen and C.-Y. Wong.


\begin{references}
\bibitem[*]{byline1} Work supported in part by the EU/TMR contract EBR
FMRX-CT96-0008.

\bibitem[**]{byline2} On leave from L.A.P.T.H.,
"Laboratoire d'Annecy-Le-Vieux de Physique Th\'eorique",  L.A.P.P.,
Chemin de Bellevue, B.P. 110, F-74941 Annecy-le-Vieux Cedex, France.

\vskip 5mm

\bibitem{rev1}
G. A. Schuler, CERN-TH.7170/94, hep-ph/9403387 and \ZP{C71}{317}{1996},
hep-ph/9504242.

\bibitem{rev2}
M. L. Mangano, Proceedings of Batavia Collider
Workshop 1995, p. 120, hep-ph/9507353.

\bibitem{rev3} A. Sansoni, Fermilab-Conf-95/263-E, Nuovo Cim.
{\bf A109}, 827 (1996) and Fermilab-Conf-96/221-E, \NP{A610}{373c}{1996}.

\bibitem{rev4} E. Braaten, S. Fleming and T. C.
Yuan, Ann. Rev. Nucl. Part. Sci. {\bf 46}, 197 (1996), hep-ph/9602374.

\bibitem{rev5} M. Beneke, Lectures given at 24th Annual SLAC Summer
Institute on Particle Physics, CERN-TH-97-055, hep-ph/9703429.

\bibitem{rev6} P. Hoyer, \NP{A622}{284c}{1997}, hep-ph/9703462.

\bibitem{rev7} M. Cacciari, DESY 97-091, hep-ph/9706374.

\bibitem{fact} J. C. Collins, D. E. Soper and G. Sterman, in
{\em Perturbative QCD}, ed. A.H. Mueller (World Scientific, 1989);
G. Bodwin, \PR{D31}{2616}{1985} and {\bf D34}, 3932 (1986) (E);
J. Qiu and G. Sterman, \NP{B353}{105}{1991} and {\bf B353}, 137 (1991).

\bibitem{masa} T. Matsui and H. Satz, \PL{178B}{416}{1986}.

\bibitem{frixione} S. Frixione, M. L. Mangano, P. Nason and G. Ridolfi,
\NP{B431}{453}{1994} and CERN-TH-97-16, hep-ph/9702287, to be
published in Heavy Flavours II, ed. by A.J. Buras and M. Lindner, World
Scientific.

\bibitem{asymhad} E.M. Aitala \etal (E791
Collaboration), \PL{B371}{157}{1996}, hep-ex/9601001.

\bibitem{asymphoto} P.L. Frabetti \etal (E687 Collaboration)
\PL{B370}{222}{1996}.

\bibitem{bmvg} S. J. Brodsky and A. H. Mueller, \PL{206B}{685}{1988};
R. Vogt and S. Gavin, \NP{B345}{104}{1990}.

\bibitem{cem1} H. Fritzsch, \PL{B67}{217}{1977}; F. Halzen,
\PL{B69}{105}{1977}.

\bibitem{cem2} R. Gavai, D. Kharzeev, H. Satz, G. A. Shuler, K. Sridhar
and R. Vogt, Int. J. Mod. Phys. {\bf A10}, 3043 (1995), hep-ph/9502270;
G. A. Schuler, CERN-TH/95-75, hep-ph/9504242;
J. A. Amundson, O. \'Eboli, E. Gregores and F. Halzen,
\PL{B372}{127}{1996}, hep-ph/9512248; and MADPH-96-942, hep-ph/9605295;
G. A. Schuler and R. Vogt, \PL{B387}{181}{1996}, hep-ph/9606410; O.
\'Eboli, E. Gregores and F. Halzen, MADPH-98-1045, hep-ph/9802421.

\bibitem{com} E. Braaten and S. Fleming, \PRL{74}{3327}{1995},
hep-ph/9411365; M. Cacciari, M. Greco, M. L. Mangano and
A. Petrelli, \PL{B356}{553}{1995}, hep-ph/9505379; P. Cho and A. K.
Leibovich, \PR{D53}{150}{1996}, hep-ph/9505329 and \PR{D53}{6203}{1996},
hep-ph/9511315.

\bibitem{nrqcd} G. T. Bodwin, E. Braaten and G. P. Lepage,
\PR{D51}{1125}{1995}, Erratum {\em ibid.,} {\bf D55}, 5853 (1997),
hep-ph/9407339.

\bibitem{csm} J. H. K\"uhn, \PL{89B}{385}{1980}; C. H. Chang,
\NP{B172}{425}{1980}; E. L. Berger and D. Jones, \PR{D23}{1521}{1981}; R.
Baier and R. R\"uckl, \PL{102B}{364}{1981} and
\ZP{C19}{251}{1983}; J. G. K\"orner, J. Cleymans, M. Karoda and G. J.
Gounaris, \NP{B204}{6}{1982}.

\bibitem{badier} J. Badier \etal (NA3 Collaboration), \ZP{C20}{101}{1983}.

\bibitem{poln} C. Biino \etal, \PRL{58}{2523}{1987};
C. Akerlof \etal (E537 Collaboration), \PR{D48}{5067}{1993};
A. Gribushin \etal (E672/E706 Collaborations) \PR{D53}{4723}{1996};
T. Alexopoulos \etal (E771 Collaboration), \PR{D55}{3927}{1997}.

\bibitem{heinrich} J. G. Heinrich \etal, \PR{D44}{1909}{1991}.

\bibitem{e705} L. Antoniazzi \etal (E705 Collaboration),
\PRL{70}{383}{1993} and \PR{D49}{543}{1994}.

\bibitem{e672} V. Koreshev \etal (E672/E706 Collaborations),
\PRL{77}{4294}{1996}.

\bibitem{vhbt} M. V\"anttinen, P. Hoyer, S. J. Brodsky and W.-K. Tang,
\PR{D51}{3332}{1995}.

\bibitem{eiqu} E. J. Eichten and C. Quigg, \PR{D52}{1726}{1995},
hep-ph/9503356.

\bibitem{bryu} E. Braaten and T. C. Yuan, \PRL{71}{1673}{1993}.

\bibitem{ic} S. J. Brodsky, P. Hoyer, C. Peterson and N. Sakai,
\PL{B93}{451}{1980}; S. J. Brodsky, C. Peterson and N. Sakai,
\PR{D23}{2745}{1981}.

\bibitem{bhmt} S. J. Brodsky, P. Hoyer, A. H. Mueller and W.-K. Tang,
\NP{B369}{519}{1992}.

\bibitem{lourenco} C. Louren\c{c}o, \NP{A610}{552c}{1996}, hep-ph/9612222.

\bibitem{knnz} B. Z. Kopeliovich and B. G. Zakharov, \PR{D44}{3466}{1991};
B. Z. Kopeliovich, J. Nemchik, N. N. Nikolaev and B. G. Zakharov,
\PL{B309}{179}{1993}, hep-ph/9305225.

\bibitem{h1ratio} C. Adloff \etal (H1 Collaboration), DESY-97-228,
hep-ex/9711012.

\bibitem{gksssv} R. Gavai, D. Kharzeev, H. Satz, G. A. Schuler, K. Sridhar
and R. Vogt, Int. J. Mod. Phys. {\bf A10}, 3043 (1995), hep-ph/9502270.

\bibitem{kzsz} M. Kr\"amer, J. Zunft, J. Steegborn and P. M. Zerwas,
\PL{B348}{657}{1995}; M. Kr\"amer, \NP{B459}{3}{1996}.

\bibitem{h1photo} S. Aid \etal (H1 Collaboration), \NP{B472}{3}{1996},
hep-ex/9603005.

\bibitem{zeusphoto} J. Breitweg \etal (ZEUS Collaboration),
\ZP{C76}{599}{1997}, hep-ex/9708010.

\bibitem{highcom}  M. Beneke, I. Z. Rothstein and M. B.
Wise, \PL{B408}{373}{1997}, hep-ph/9705286; B. A. Kniehl and G. Kramer,
DESY 98-023, hep-ph/9803256.

\bibitem{dprod} G. A. Alves \etal, \PRL{70}{722}{1993}; M. J. Leitch
\etal, \PRL{72}{2542}{1994}.

\bibitem{klns} D. Kharzeev, C. Louren\c{c}o, M. Nardi and H. Satz,
\ZP{C74}{307}{1997}.

\bibitem{kopeliovich} B. G. Kopeliovich, Proc. Hirschegg 1997, {\em QCD
phase transitions}, p. 281, hep-ph/9702365.

\bibitem{kharzeev} D. Kharzeev, talk at the Quark Matter '97 Conference,
nucl-th /9802037.

\bibitem{khasat} D. Kharzeev and H. Satz, \PL{B366}{316}{1996}.

\bibitem{nmc} P. Amaudruz \etal (NMC Collaboration), \NP{B371}{553}{1992}.

\bibitem{goupir} T. Gousset and H. J. Pirner, \PL{B375}{349}{1996}.

\bibitem{abe} F. Abe \etal (CDF Collaboration), \PRL{75}{4358}{1995};
T. Alexopoulos \etal (E771 Collaboration), \PL{B374}{271}{1996}.

\bibitem{alde} D. M. Alde \etal (E772 Collaboration),
\PRL{66}{2285}{1991};
M. J. Leitch \etal (E789/E772 Collaborations), \NP{A544}{197c}{1992}.

\bibitem{benrot}  M. Beneke and I.Z. Rothstein, \PR{D54}{2005}{1996},
Erratum \PR{D54}{7082}{1996}, hep-ph/9603400.

\bibitem{cholei}  P. Cho and A. K. Leibovich, \PR{D53}{150}{1996},
hep-ph/9505329 and \PR{D53}{6203}{1996}, hep-ph/9511315.

\bibitem{isgurwise} N. Isgur and M. B. Wise, \PL{B232}{113}{1989}.

\bibitem{hope} P. Hoyer and S. Peign\'e, \PR {D57}{1864}{1998},
hep-ph/9706486.

\bibitem{pdg} R. M. Barnett \etal (Particle Data Group), \PR{D54}{1}{1996}.

\bibitem{spinproj} J. H. K\"uhn, J. Kaplan and E. G. O. Safiani,
\NP{B157}{125}{1979}.

\bibitem{tava} W.-K. Tang and M. V\"anttinen, \PR{D53}{4851}{1996} and
\PR{D54}{4349}{1996}.

\end{references}
\end{document}